\documentclass[
 reprint,
 superscriptaddress,
 amsmath,amssymb,
 aps,
 prb,
]{revtex4-2}

\usepackage{graphicx}
\usepackage{dcolumn}
\usepackage{bm}

\begin{document}

%\preprint{APS/123-QED}

\title{X-ray ptychographic topography, a new tool for strain imaging}

\author{Mariana Verezhak}
\email[]{Corresponding author: mariana.verezhak@psi.ch}
\affiliation{Paul Scherrer Institute, Forschungsstrasse 111, 5232 Villigen PSI, Switzerland}

\author{Steven Van Petegem}
\affiliation{Paul Scherrer Institute, Forschungsstrasse 111, 5232 Villigen PSI, Switzerland}

\author{Angel Rodriguez-Fernandez}
\affiliation{European XFEL, Holzkoppel 4, 22869 Schenefeld, Germany}

\author{Pierre Godard}
\affiliation{Institut Pprime, CNRS-University of Poitiers-ENSMA, SP2MI, Futuroscope, France}

\author{Klaus Wakonig}
\affiliation{Paul Scherrer Institute, Forschungsstrasse 111, 5232 Villigen PSI, Switzerland}
\affiliation{ETH and University of Zürich, Institute for Biomedical Engineering, 8093 Zürich, Switzerland}

\author{Dmitry Karpov}
\affiliation{Paul Scherrer Institute, Forschungsstrasse 111, 5232 Villigen PSI, Switzerland}

\author{Vincent L.R. Jacques}
\affiliation{Laboratoire de Physique des Solides UMR8502, CNRS, Université Paris-Saclay, 91405, Orsay, France}

\author{Andreas Menzel}
\affiliation{Paul Scherrer Institute, Forschungsstrasse 111, 5232 Villigen PSI, Switzerland}

\author{Ludovic Thilly}
\affiliation{Institut Pprime, CNRS-University of Poitiers-ENSMA, SP2MI, Futuroscope, France}

\author{Ana Diaz}
\affiliation{Paul Scherrer Institute, Forschungsstrasse 111, 5232 Villigen PSI, Switzerland}

\date{\today}

\begin{abstract}
Strain and defects in crystalline materials are responsible for the distinct mechanical, electric and magnetic properties of a desired material, making their study an essential task in material characterization, fabrication and design. Existing techniques for the visualization of strain fields, such as transmission electron microscopy and diffraction, are destructive and limited to thin slices of the materials. On the other hand, non-destructive X-ray imaging methods either have a reduced resolution or are not robust enough for a broad range of applications. Here we present X-ray ptychographic topography, a new method for strain imaging, and demonstrate its use on an InSb micro-pillar after micro-compression, where the strained region is visualized with a spatial resolution of 30 nm. Thereby, X-ray ptychographic topography proves itself as a robust non-destructive approach for the imaging of strain fields within bulk crystalline specimens with a spatial resolution of a few tens of nanometers.
\begin{description}
\item[Subject Areas]
Materials Science, Interdisciplinary Physics, Nanophysics. 
\end{description}
\end{abstract}

\maketitle

\section{Introduction}

Progress in advanced materials, from lightweight composites to biomedical technology, relies on understanding the relationship between the structure and properties of a material. In addition to the chemical composition and crystalline lattice, strain fields caused by defects also regulate material properties. Therefore, the detection and characterization of strain and its relation to the type and concentration of defects in the crystal at the nanoscale is an essential, albeit challenging task.

\begin{figure*}[t]
\includegraphics[width=1.\textwidth]{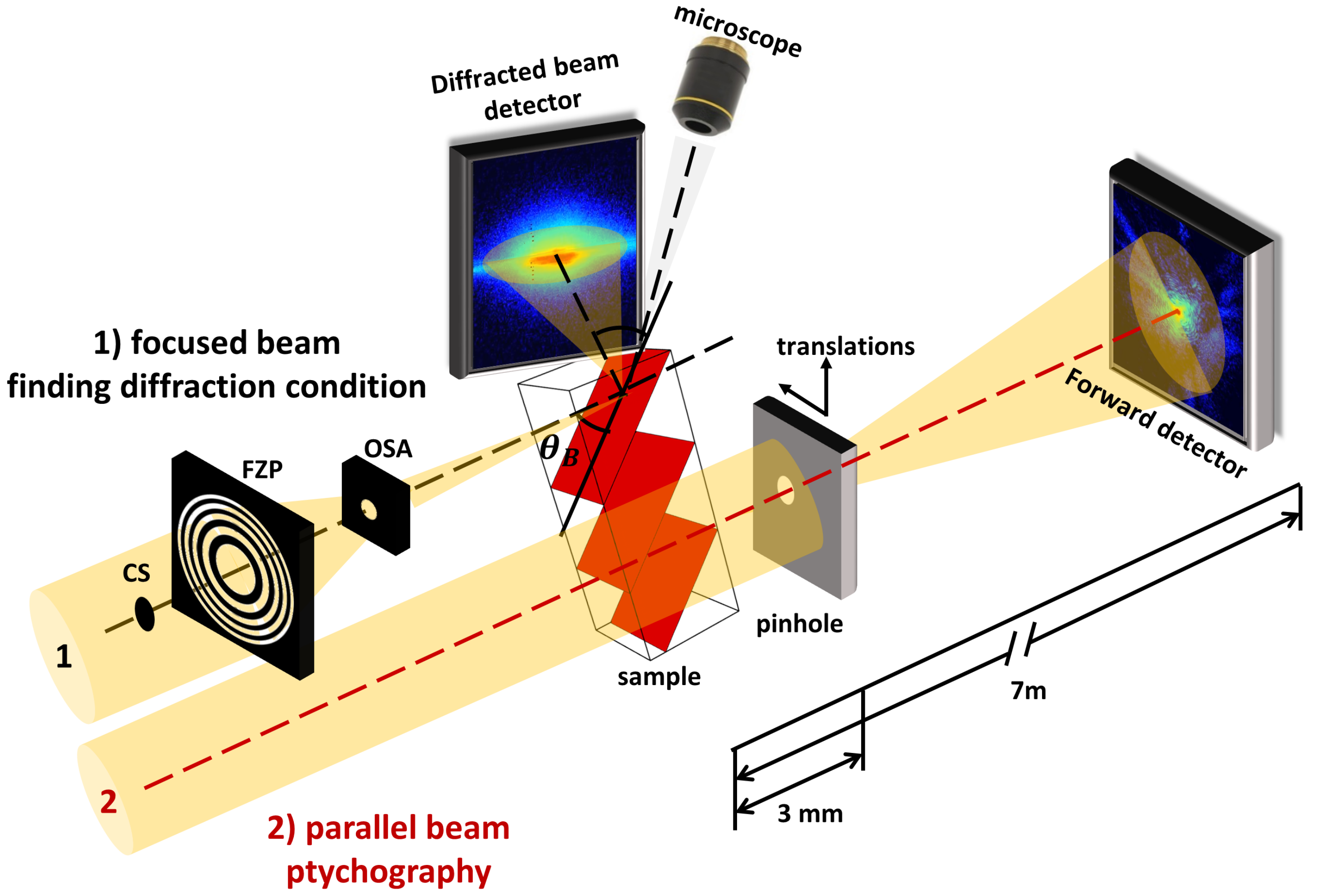}
\caption{\label{fig:f1}Ptychographic topography acquisition scheme. Step 1) focused beam: sample alignment and diffraction peak detection from non-strained crystal region using microscope, focusing optics and 2D detector along the diffracted beam. Focusing optics includes Fresnel zone plate (FZP), central stop (CS) and order-sorting aperture (OSA). Step 2) parallel beam: ptychographic topography with parallel coherent beam.  A pinhole is scanned after the sample and far-field patterns are recorded with the detector along the forward direction. Note: the two steps are performed consecutively.}
\end{figure*}

Transmission electron microscopy (TEM) provides imaging of crystalline defects with atomic spatial resolution. However, due to the limited penetration depth of electrons, invasive sample preparation of thin sections is required, which can modify the strain fields to be analyzed \cite{1Hytch2014}. X-ray diffraction, on the other hand, is a non-invasive alternative which is sensitive to atomic displacements \cite{2Hrauda2011}. X-ray topography (XRT) \cite{3Ramachandran1944} has been routinely used for imaging the defect microstructure in the micrometer- to centimeter-sized crystals based on the diffraction contrast. In XRT, the image contrast comes from variations in the crystal lattice spacing and/or orientation, with the resolution limited by the detector pixel size.

To achieve higher spatial resolution with X-rays, X-ray focusing optics were introduced, as for example in Laue X-ray micro-diffraction \cite{4Ice2009}. Here, the diffraction patterns are obtained with a polychromatic X-ray beam and each diffraction spot comes from a different crystalline plane. The peak position depends on the average crystal orientation and unit cell parameters, while the width provides information on the strain gradient within the illuminated volume. Therefore, any change in peak positions and/or widths is the footprint of a specific strain field in the crystal caused by the presence of particular defects or lattice distortions. In such experiments, spatial resolution is achieved by focusing the beam and raster scanning the sample at different positions \cite{5Maas2009}. Monochromatic X-ray diffraction microscopy in full field \cite{6Simons2015} and scanning mode \cite{7Etzelstorfer2014}, \cite{8Chahine2014} have also been able to reveal the strain field in crystalline samples with the spatial resolution limited by the focusing lens.

With advancements in synchrotron brilliance and degree of spatial coherence, probing individual crystalline defects like dislocations became possible \cite{9Jacques2011}. Coherent X-ray diffraction microscopy has enabled nanometer spatial resolution beyond focusing optics limitations \cite{10Miao2015}, \cite{11Pfeiffer2018}. Bragg coherent diffraction imaging (BCDI) is based on measuring the far-field diffraction patterns of a fully illuminated crystal placed in the Bragg condition and subsequent application of a phase retrieval algorithm for image reconstruction. Thus, BCDI provides 3D maps of the atomic displacement field within a small, isolated nanocrystal with a spatial resolution of a few tens of nanometers \cite{12Pfeifer2006}. On the other hand, ptychographic methods, based on measuring a series of diffraction patterns at a set of overlapping areas \cite{13Rodenburg2007}, enable the application of coherent methods to larger crystalline specimens. Bragg ptychography is also capable of imaging strain; however, the strict requirements on the specimen translation and beam stability make the technique difficult to apply, especially in 3D \cite{14Godard2011a}. Bragg projection ptychography \cite{15Hruszkewycz2017} has more relaxed stability requirements but due to the divergence of the incoming beam the method is not working for crystals with lattice displacements within the scattering plane. In such cases, it is not always the same part of the convergent incoming beam that is diffracted, contradicting the hypothesis of a constant probe, as discussed in \cite{16Tsai2016}. This reduces the robustness of the image reconstruction in Bragg ptychography.

Taking advantage simultaneously of the high strain sensitivity, large field of view, and simplicity of X-ray topography, as well as the high spatial resolution of X-ray ptychography, we propose a new X-ray imaging technique, ptychographic topography, for robust and flexible strain field imaging in extended crystalline samples with high spatial resolution. This method exploits the high strain sensitivity provided by a parallel, monochromatic and coherent incident beam in combination with the concept of tele-ptychography \cite{16Tsai2016} to reconstruct the exit wave field a few millimeters downstream of the sample. Subsequent numerical backpropagation of the reconstructed wave field to the sample position provides the exit wave after parallel X-rays propagate through the crystal. This new approach offers straightforward reconstructions, even in the case of complicated strain fields, with the resolution not limited by lenses.

In this paper, we describe the application of this technique to strain mapping of InSb micro-pillars that were uniaxially compressed up to the plastic deformation regime. The strain field is visualized in two different configurations of ptychographic topography, in forward and diffraction directions. The advantages and limitations of both configurations are explained and the effect of dynamical diffraction on the observed results is discussed.

\begin{figure*}[t]
\includegraphics[width=1.\textwidth]{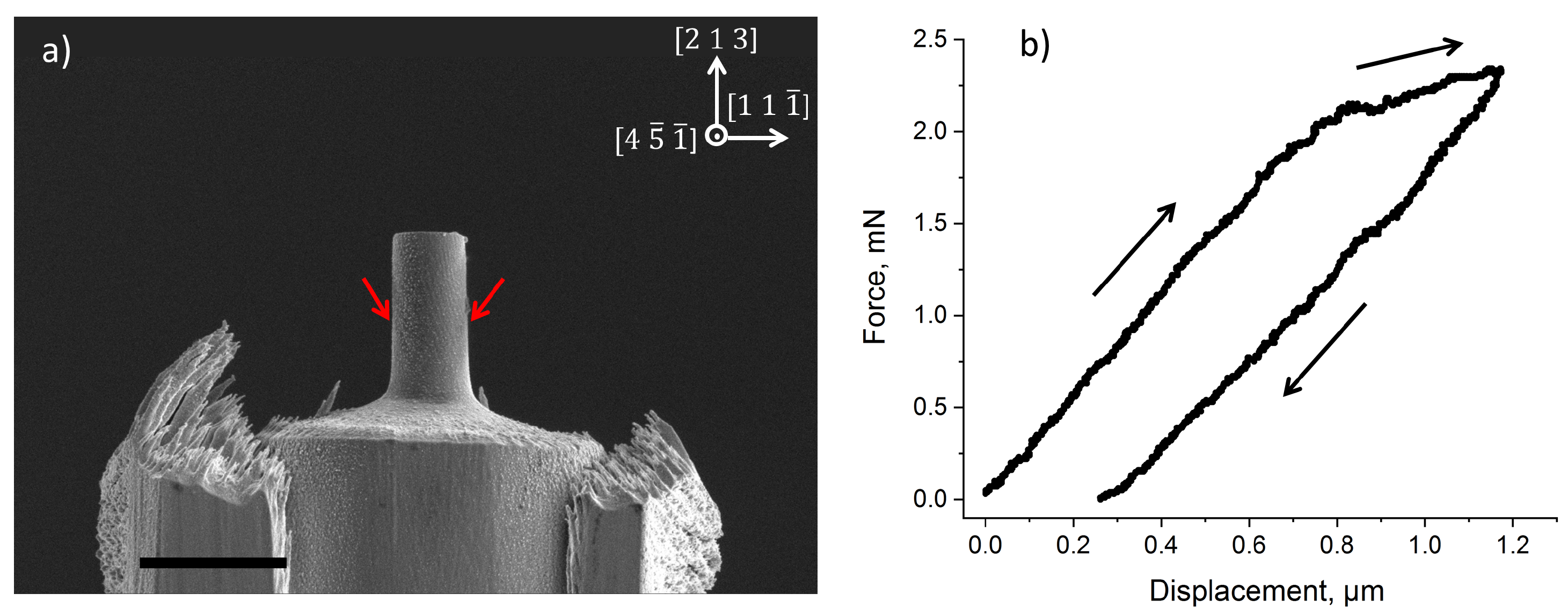}
\caption{\label{fig:f2}Mechanical deformation of InSb micro-pillars. (a) SEM image of the pillar and its pedestal after uniaxial compression along the [2 1 3] crystallographic direction to reach the plastic deformation regime (applied force of 2.35 mN). Scale bar: 5 µm.  (b) Force-displacement curve of the compression test that confirms that the sample was plastically deformed.}
\end{figure*}

\section{Methods}

\subsection{The principle of ptychographic topography}

The concept of ptychographic topography consists of placing the crystal in diffraction condition for specific crystallographic planes and performing ptychographic scans in forward (or diffraction) direction with a pinhole scanned several millimeters after the sample to reconstruct the exit wave field at the pinhole position. The resulting image is then numerically back propagated to the crystal position and is sensitive to lattice imperfections.

As shown in Fig.~\ref{fig:f1}, first, the sample's region of interest is selected via an auxiliary visible light microscope. While setting up the experiment, we used a focused beam configuration to illuminate a region of the sample which has no crystalline defects, and to find and center the diffraction peak on the detector placed along the diffracted beam as shown in the step 1 in Fig.~\ref{fig:f1}. 

To perform the ptychographic topography measurement, we switch to the parallel beam configuration (step~2 in Fig.~\ref{fig:f1}). A pinhole is placed a few millimeters after the sample and is scanned perpendicular to the beam, providing the sufficient overlap of the pinhole area at neighboring scan positions required for ptychographic reconstructions \cite{16Tsai2016}, \cite{17Verezhak2018}. The diffraction patterns are recorded with a 2D detector downstream of the pinhole along the forward (or diffraction) direction in the far field and are used for the reconstruction of the wavefront at the pinhole position using ptychographic phase retrieval. Numerical backpropagation to the sample plane then results in an image of the sample sensitive to the lattice displacements caused by defects. 

By rotating the sample and performing measurements at different angular steps close to a diffraction condition, quantitative information about the amount of strain and lattice rotations can be obtained. To recover all components of the three-dimensional displacement field one would need to measure at several non-equivalent diffraction peaks.

\subsection{InSb micro-pillars upon uniaxial compression}

Focused ion beam (FIB) milled InSb single-crystalline cylindrical micro-pillars, 2 µm in diameter and 6 µm in height on cylindrical pedestals, were used to demonstrate the potential of the method. A scanning electron microscopy (SEM) image of such a pillar is shown in Fig.~\ref{fig:f2}(a). Two equivalent pillars (S1 and S2) were used for ptychographic topography in order to show reproducibility. The results from sample S1 are presented below while those from sample S2 are shown in the Supplemental Material.

The micro-pillars were uniaxially compressed with a dedicated micro-compression device \cite{35Kirchlechner2011} along the pillar vertical axis that corresponds to the [2 1 3] direction, favoring only one dislocation slip system \cite{18Thilly2012}. The pillars were compressed up to a force of 2.35 mN, corresponding to the plastic regime, which led to an irreversible reduction of the pillars height of about 300 nm (i.e. about 5\% of their initial height). This can be seen as a residual displacement after unloading at the force-displacement curve shown in Fig.~\ref{fig:f2}(b). In addition, slip traces (pointed to with red arrows in Fig.~\ref{fig:f2}(a)) are visible at the pillar surface. More information about sample preparation can be found in the Supplemental Material.

\section{Ptychographic topography in forward and diffraction geometry}

\begin{figure*}[t]
\includegraphics[width=1.\textwidth]{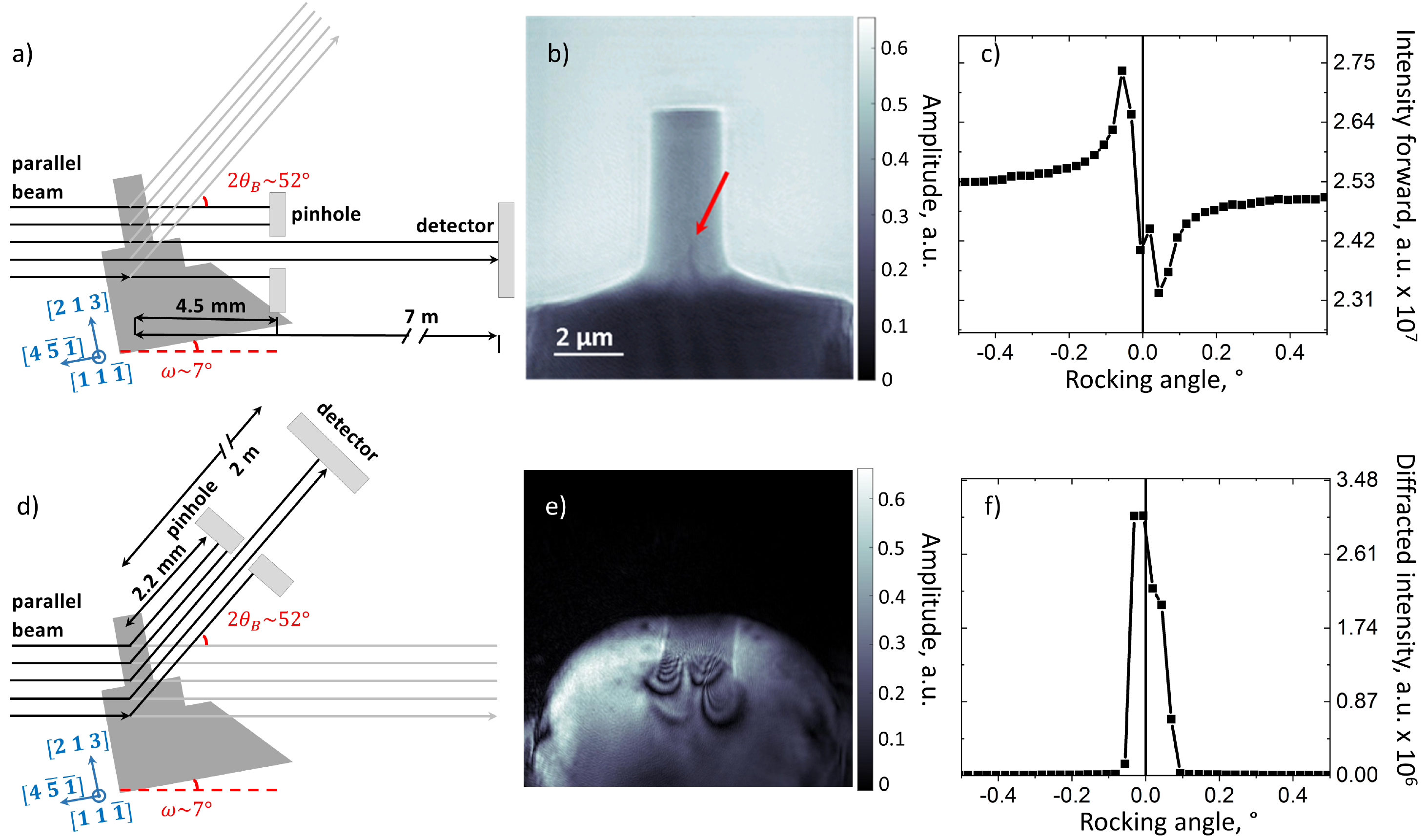}
\caption{\label{fig:f3}Ptychographic topography of InSb micro-pillars in forward and diffraction directions. Schemes of ptychographic topography in (a) forward and (d) diffraction direction with sample orientation (sample not to scale). (b) Amplitude of the image reconstruction obtained in forward direction at the rocking angle of 0.03°. Red arrow shows strained region on the image. (c) Forward direction rocking curve around  the (2 0 2) diffraction peak measured with focused beam at the pedestal position.  (e) Amplitude of the image reconstruction in diffraction direction at the rocking angle of 0.002°. (f) Rocking curve in diffraction direction around the (2 0 2) diffraction peak measured with focused beam at the pedestal position.}
\end{figure*}

We demonstrate the implementation of ptychographic topography in both forward and diffraction direction at the cSAXS beamline at the Swiss Light Source (SLS). In both cases, we first used the focused beam configuration with an illumination spot size of $\sim$100 nm in diameter. InSb samples were then brought to the (2 0 2) diffraction condition (similar to the geometry in \cite{19Jacques2013}) as shown in Fig.~\ref{fig:f3}(a) and Fig.~\ref{fig:f3}(d). 

The region well below the pillar, i.e. the pedestal, was chosen for the diffraction peak alignment as a region without crystalline defects. The rocking curves were recorded with detectors placed both along the forward and the diffraction direction. The rocking curves obtained at the pedestal position are shown in  Fig.~\ref{fig:f3}(c) and  Fig.~\ref{fig:f3}(f). More details about experimental alignment are presented in the Supplemental Material.

For forward ptychographic topography, the setup was then switched to the parallel-beam configuration by removing focusing optics and placing a pinhole of $\sim$3.5 µm diameter 4.5 mm downstream the sample. Samples were rocked around the (2 0 2) diffraction peak while, at each rocking angle, the pinhole was scanned after the sample perpendicular to the incoming beam. The schematic of the described experiment is shown in  Fig.~\ref{fig:f3}(a). For more details see the Supplemental Material. 

The second experiment was performed in the diffraction geometry, see schematic in  Fig.~\ref{fig:f3}(d) and the Supplemental Material. In this case, the 2 µm pinhole was placed at 2.2 mm downstream the sample in the diffracted beam direction and used for scanning. Samples were rocked around  the (2 0 2) diffraction peak and a ptychographic scan was performed at each rocking angle.

Ptychographic reconstructions were performed using 300 iterations of the difference map algorithm \cite{37Thibault2009} followed by 1000 iterations of a maximum likelihood refinement \cite{38Thibault2012} using the PtychoShelves package \cite{39Wakonig2020}. The resulting ptychographic reconstructions with a pixel size of 21 nm (in forward geometry) and 30 nm (in diffraction geometry) were back-propagated to the sample position and followed by post-processing (phase offset and a linear phase wrap removal, vertical and horizontal spatial alignment) \cite{40Guizar-Sicairos2011}. We estimated the resulting spatial resolution using Fourier ring correlation \cite{25VanHeel2005} to be 29 nm in the forward direction (see Fig. S2 in the Supplemental Material). 

\section{Strain and dynamical effects in InSb}

As mentioned above, the 2D ptychographic projections were taken at different angular positions in the vicinity of the InSb (2 0 2) diffraction peak. This reflection is in Laue diffraction geometry, in which the diffracted wave exits the crystal at the opposite surface to the incoming beam \cite{20Rodriguez-Fernandez2018}. The rocking curve was acquired with a focused beam in the forward direction at the pedestal position where we expect only negligible strain (Fig.~\ref{fig:f3}(c)). The curve shows increased intensity at angles just below the diffraction condition and exhibits an intensity drop at the diffraction angle. Such effect shows that dynamical scattering effects take place in this region of the crystal. This is expected as the pedestal is 10 µm thick and the extinction length of InSb at 6.2 keV and  (2 0 2) diffraction reflection is 4.98 µm, as estimated from \cite{21Authier2010}. Such effect is well-known as a dynamical diffraction effect \cite{22Zachariasen1968}, \cite{23Batterman1964} and it highlights the good crystallinity at the pedestal far away from the strained micro-pillar.

\begin{figure}[t]
\includegraphics[width=0.5\textwidth]{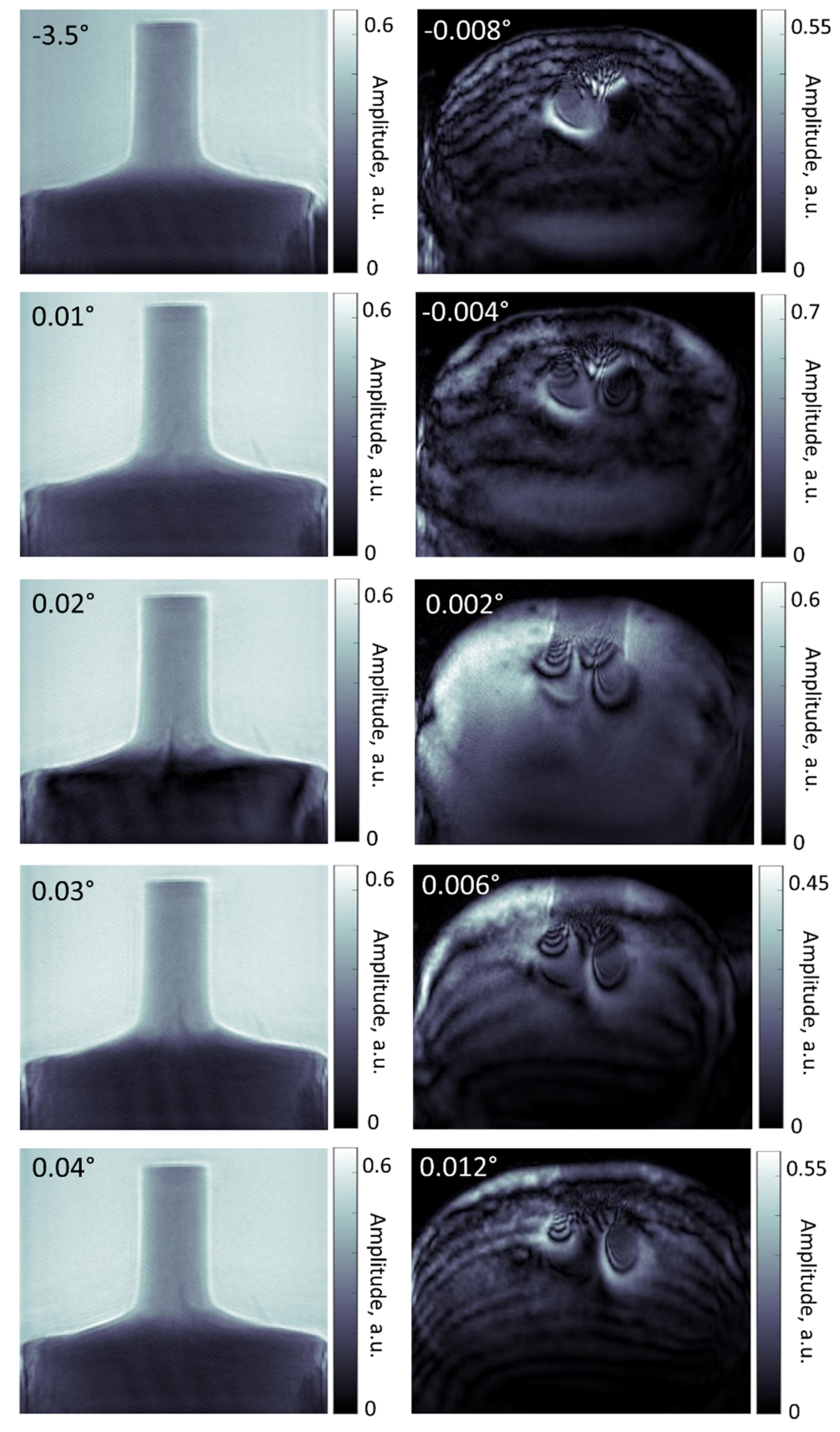}
\caption{\label{fig:f4}Evolution of ptychographic topography of InSb micro-pillars in forward and diffraction geometry with the rocking angle around the (2 0 2) reflection. The rocking angle is shown in the top left corner of individual images. Left column: amplitudes of the image reconstruction obtained in forward direction. The image at the rocking angle of -3.5° is acquired far away from the diffraction condition and shows no strain features. Other images in vicinity of the diffraction condition show the strain feature. Right column: amplitudes of the image reconstruction obtained in diffraction direction, showing circular features and fringes of varying contrast.}
\end{figure}

Each projection of the ptychographic topography experiment in forward direction was reconstructed using single-probe mode \cite{24Thibault2013} in order to reconstruct the pinhole,  see Fig. S1(a). The amplitude (Fig.~\ref{fig:f3}(b)) of the ptychographic reconstruction clearly shows the pillar and pedestal shapes. At the bottom of the pillar and at the pillar/pedestal interface, a dark feature is visible, indicated by the red arrow in Fig.~\ref{fig:f3}(b). This feature is only visible in the vicinity of the diffraction peak angle (as shown in Fig.~\ref{fig:f4} for rocking angles 0.01-0.04°) and is not observed far away from the diffraction peak (as shown in Fig.~\ref{fig:f4} for a rocking angle of -3.5°). All projections can also be accessed in the Media S1. This feature corresponds to crystallinity imperfection or strain field present in the crystal that makes this part of the crystal diffract at slightly different angle. The resolution of the reconstructed amplitude image was estimated to be 29 nm by Fourier ring correlation \cite{25VanHeel2005}, as shown in Fig. S2.

It is also important to note that in ptychographic topography the amplitude provides higher contrast of crystalline imperfections compared to the phase reconstruction, as can be seen in Fig. S3 in the Supplemental Material. This can be understood as a result of the diffraction-based contrast, similar to the one in X-ray topography. Nevertheless, we note that the phase information is needed for the back propagation of the exit wave field from the pinhole to the sample position.

In the case of ptychographic topography in diffraction direction, the reconstruction improved taking into account a second incoherent mode, with an intensity distribution for the first and second modes of 77\% and 23\%, respectively (see in Fig. S1(b) and Fig. S1(c)). This is similar to what is observed with Bragg ptychography as the reconstructions are more difficult compared to conventional ptychography and sometimes require several incoherent modes. As the signal in the diffraction direction is often several orders of magnitude smaller than in the direct transmitted beam, there is less data available while the system might also have higher stability requirements, which affects the robustness of the reconstruction.

The resulting reconstruction along the diffraction direction (Fig.~\ref{fig:f3}(e)) contains no information at the top half of the images (black region in Fig.~\ref{fig:f3}(e)). This is due to the diffraction angle $\bm{2 \theta_B\sim52^{\circ}}$ at which those projections were recorded. At this angle, we see a projection of the whole pillar/pedestal assembly along the diffracted beam direction. In order to illustrate this effect, we calculated the projected thickness of the sample, including pillar and pedestal, along both forward and diffraction direction in our measurement geometry (see Fig. S4 in the Supplemental Material). The projected thickness of the sample, shown in Fig. S4(d) has a similar shape as observed in the image reconstructions along the diffraction direction and shows that the information of the whole pillar as well as the pedestal is encoded in the bottom half of the reconstructed image.

A very interesting feature can be seen in the ptychographic topography reconstructions obtained in diffraction geometry (see Fig.~\ref{fig:f3}(e)): two sets of circular fringes of varying contrast. These features strongly change with the rocking angle and more interferences appear as we go further away from the diffraction angle, as shown in Fig.~\ref{fig:f4}, right column. The projections at all rocking angles obtained by ptychographic topography in diffraction direction can be seen in Media S2.

The interpretation of these features is complicated as it can involve the simultaneous presence of different effects. First of all, the thickness variation of single-crystals can cause so-called Pendellösung fringes in the presence of dynamical scattering, i.e. if the crystal thickness is larger than the extinction length. This effect is widely observed in X-ray topography caused by thickness variations \cite{26Kato1959}. We believe that the thickness-based Pendellösung effect does not cause the observed fringes at the diffraction direction (Fig.~\ref{fig:f3}(e)) as the projected thickness variation in the diffraction direction does not show similar patterns or significant sudden thickness change at the base of the pillar. However, thickness-based Pendellösung effect can very well explain the appearance and the shape of the additional fringes in the region of the pedestal as we go further away from the diffraction peak, as shown in Fig.~\ref{fig:f4}, right column. 

To confirm our hypothesis, we qualitatively simulated the absorption signal from the projected thicknesses using pencil-beam symmetrical Laue dynamical theory based on the Eq. 16 in Ref. \cite{27Punegov2016}. At some rocking angles, we were able to observe similarities between these fringes that appear further away from the diffraction condition with the simulated ones (reported in the Supplemental Material, Fig. S5). More detailed simulation, taking into account the non-symmetric Laue dynamical theory as well as non-pencil beam is required to further understand this phenomenon but is beyond the scope of this paper.

A second possible effect, which could not be properly taken into account in our simulations, is the possibility of additional destructive and constructive interferences that emerge due to the beam path differences caused by our large non-focused incident beam, so called Borrmann-Lehmann interference patterns \cite{28Mai1989}, \cite{29Lang1990}. This effect, which is also caused by dynamical scattering, could very well contribute to the observed fringes.

Finally, strain can also cause interference fringes. This effect is extensively studied with X-ray topography using different single crystalline specimens with different strain states \cite{30Bonse1969, 31Negishi2007, 32Chukhovskii1988, 33Novikov2010}. We believe that the fringes we observe with ptychographic topography in diffraction direction are mainly caused by the strain present in the crystal in combination with the interference effects caused by the shape, thickness and orientation of our sample. In addition, any interference fringes caused by sample size and shape are extremely sensitive to strain. The fringes symmetry, shape and number will change in the presence of characteristic strain or lattice imperfections \cite{29Lang1990}. This makes ptychographic topography in diffraction direction extremely sensitive to small crystalline imperfections and strain fields.

The stress-displacement curve in Fig.~\ref{fig:f2}(b) shows that there has been a 5\% irreversible deformation in the total height of the pillar, which is a sign of considerable plastic deformation in the pillar. This was confirmed by SEM images performed after the compression experiment, like that shown in Fig.~\ref{fig:f2}(a), which reveal the presence of slip traces, a typical signature of plastic deformation in this system \cite{18Thilly2012}. Furthermore, a large amount of strain most probably in the form of lattice rotations is produced in addition to the slip planes. This was confirmed by performing angular rocking scans around the (2 0 2) diffraction condition with the focused beam at different positions along the pillar and recording the intensity with the 2D detector along the diffraction direction, see Fig. S6. The rocking curve at the pedestal is narrow, as expected for a perfect crystal, and the curve becomes broader and significantly shifted as the beam impinges higher on the pillar. Additionally, scanning diffraction microscopy performed with the focused beam at a fixed angle close to the (2 0 2) diffraction condition also revealed strained areas within the pillar and at the pillar/pedestal interface, as shown in Fig. S7. 

\section{Discussion}

In the ptychographic topography demonstration described here, we observe deviations from perfect crystallinity due to strain and lattice rotations especially at the bottom part of the pillar, both along the forward and the diffraction directions. With a parallel incident beam, it is clear that a certain degree of crystallinity within the sample is needed in order to observe deviations from that crystallinity, and this is the case at the bottom part of the pillar. On the other hand, on the top part of the pillar the strain affects most of the bulk of the sample, in such a way that only a small sample volume diffracts at a given angle. Hence, for heavily strained crystals, the diffraction signal is too weak to be reconstructed in the diffraction geometry, and the diffraction-based contrast is too low in the forward geometry. Nevertheless, the advantage of having a parallel beam is that the technique can potentially have a very high sensitivity to strain in a crystalline specimen. As the contrast is the same as in X-ray topography, we expect that the strain sensitivity could be as good as $\bm{\Delta d/d=1e^{-7}}$, where $\bm{d}$ is the lattice spacing of the crystal and $\bm{\Delta d}$ is a change in $\bm{d}$ over the initial length.

For illustration of reproducibility, the same ptychographic topography measurement was conducted along the (2 0 2) diffraction direction on a similar pillar, sample S2, that had also been previously mechanically compressed, see Fig. S8. The reconstructed amplitudes along the diffraction direction show many similarities in the type, contrast and spacing of the fringes, although their exact arrangement is not identical and could be explained by different strain and/or lattice rotation distribution inside the pillar. Sample S2 was also investigated by ptychographic topography at  the (2 2 0) Bragg condition along the forward direction (see Fig. S9), revealing at some angles a crystalline imperfection feature (Fig.~S9(b)) similar to that observed in sample~S1 at  the~(2~0~2) diffraction condition (Fig.~\ref{fig:f3}(b)). At a different angle, other interesting features were also observed (Fig. S9(c)). However, the exact determination of the type of defects seen here will require further studies.

In this paper, we report a new robust strain imaging technique, ptychographic topography, as well as its realization in both forward and diffraction direction. This new approach allows visualization of the exit wave after propagation of a beam through the crystal with high resolution (at least 29 nm as achieved in the present work) and both absorption and phase contrast. In the forward direction, we obtain a superposition of the strained areas within the sample and the transmission image of the specimen, which is useful to study strain in a crystal in context with its surrounding material, regardless of its structural nature. In the diffraction direction, we observed fringes caused by dynamical scattering effects due to strain and geometrical features in the sample both in absorption and phase contrast. 

The use of a parallel beam offers opportunities for very high strain sensitivity when used on samples that present a sufficient degree of crystallinity. Signal-to-noise in ptychographic topography could be improved for such highly strained samples with the use of a coherently illuminated condenser designed to produce a flat illumination for X-ray transmission  microscopes \cite{34Jefimovs2008}. With the currently available coherent flux in 3rd generation synchrotron sources like the SLS, we estimate that coherent flux could be improved by about a factor of 10 already now by using this technology. With upcoming upgrades of synchrotron storage rings, an additional significant increase in the coherent flux will be possible.

\begin{acknowledgments}
We acknowledge the Paul Scherrer Institute, Villigen, Switzerland for provision of synchrotron radiation beamtime at the cSAXS beamline of the SLS. This work was supported by the SNF grant No 200021L\_169753, ANR-16-CE93-0006, by “Investissement d'Avenir” (LABEX INTERACTIFS, ANR-11-LABX-0017-01) and by Nouvelle Aquitaine Region / European Structural and Investment Funds (ERDF No P-2016-BAFE-94/95). M.V. was supported by the European Union's Horizon 2020 research and innovation program under the Marie Skłodowska-Curie grant agreement No 701647. K.W. acknowledges the support by the SNF grant  No~200021\_166304. Authors thank M. Guizar-Sicairos, P.~O.~Renault, T. Sadat, D. Le Bolloc'h, B. Kedjar and F.~Mignerot for fruitful discussions.
\end{acknowledgments}

%\bibliography{biblio}

\providecommand{\noopsort}[1]{}\providecommand{\singleletter}[1]{#1}%

\end{document}

%\preprint{APS/123-QED}

\title{Supplemental Material for: X-ray ptychographic topography, a new tool for strain imaging}

\author{Mariana Verezhak}
\email[]{Corresponding author: mariana.verezhak@psi.ch}
\affiliation{Paul Scherrer Institute, Forschungsstrasse 111, 5232 Villigen PSI, Switzerland}

\author{Steven Van Petegem}
\affiliation{Paul Scherrer Institute, Forschungsstrasse 111, 5232 Villigen PSI, Switzerland}

\author{Angel Rodriguez-Fernandez}
\affiliation{European XFEL, Holzkoppel 4, 22869 Schenefeld, Germany}

\author{Pierre Godard}
\affiliation{Institut Pprime, CNRS-University of Poitiers-ENSMA, SP2MI, Futuroscope, France}

\author{Klaus Wakonig}
\affiliation{Paul Scherrer Institute, Forschungsstrasse 111, 5232 Villigen PSI, Switzerland}
\affiliation{ETH and University of Zürich, Institute for Biomedical Engineering, 8093 Zürich, Switzerland}

\author{Dmitry Karpov}
\affiliation{Paul Scherrer Institute, Forschungsstrasse 111, 5232 Villigen PSI, Switzerland}

\author{Vincent L.R. Jacques}
\affiliation{Laboratoire de Physique des Solides UMR8502, CNRS, Université Paris-Saclay, 91405, Orsay, France}

\author{Andreas Menzel}
\affiliation{Paul Scherrer Institute, Forschungsstrasse 111, 5232 Villigen PSI, Switzerland}

\author{Ludovic Thilly}
\affiliation{Institut Pprime, CNRS-University of Poitiers-ENSMA, SP2MI, Futuroscope, France}

\author{Ana Diaz}
\affiliation{Paul Scherrer Institute, Forschungsstrasse 111, 5232 Villigen PSI, Switzerland}

\date{\today}

\maketitle

\section{Media}

Media S1. Ptychographic topography of InSb micro-pillar in forward direction at different angular positions of the rocking curve around the (2 0 2) diffraction peak. 

Media S2. Ptychographic topography of InSb micro-pillar in diffraction direction at different angular positions of the rocking curve around the (2~0~2) diffraction peak.

\section{Experimental details}

The samples were prepared by ion milling with Ga+ focused ion beam (FIB) on an InSb single-crystalline wedge with known orientation in the form of a series of cylindrically shaped micro-pillars on cylindrical pedestals. The pillars dimensions are 2 $\mu$m in diameter and 6 $\mu$m in height. The pedestals dimensions are 10 $\mu$m in diameter and 10 $\mu$m in height. Two equivalent pillars (S1 and S2) were investigated by ptychographic topography for reproducibility. The results from pillar S1 are shown in the main text while those from pillar S2 in the Supplemental Material.

The samples were uniaxially compressed with a dedicated micro-compression device equipped with a diamond flat punch \cite{35Kirchlechner2011}. The compression axis was chosen along to the [2 1 3] direction that corresponds to the pillar vertical axis. The pillars were compressed up to a force of 2.35 mN and the compression tip was retracted. 

The ptychographic topography setup was built at the cSAXS beamline at the SLS. For the focused beam configuration, we used a Fresnel zone plate of 200 $\mu$m diameter and 90 nm outermost zone width upstream of the sample with a central stop of 60 $\mu$m diameter and an order-sorting aperture of 30 $\mu$m diameter. Such optical layout allowed achieving an illumination spot size of $\sim$100 nm in diameter. InSb samples were then brought to the (2 0 2) diffraction condition at 6.2 keV, corresponding to a Bragg angle  $\bm{\theta_B=25.88^{\circ}}$ and an orientation angle of the sample $\bm{\omega \sim 7^{\circ}}$, as shown in Fig.3(a) and Fig.3(d) (main text). The diffraction peak alignment was performed with the beam at the pedestal position. The rocking curves were recorded with an Eiger 500K detector placed along the diffraction direction and a \mbox{Pilatus 2M} detector along the forward direction. 

For ptychographic topography in forward geometry we used the parallel-beam configuration, as shown in Fig.1 (main text). A pinhole of 3.5 $\mu$m diameter was placed on a 2D piezo translation stage 4.5 mm downstream the sample in forward direction. Samples were rocked around the (2 0 2) diffraction peak in a range of $\pm$0.7$^{\circ}$ with a step size of 0.01$^{\circ}$. At each rocking angle, the pinhole was scanned after the sample perpendicular to the incoming beam with a step size of 0.5 $\mu$m and 0.2 s exposure time covering a field of view of 12 x 12 $\mu$m$^2$. The scanning positions followed a Fermat spiral pattern \cite{36Huang2014}. The diffraction patterns were recorded with a Pilatus 2M detector, placed at 7.338 m from the sample in the forward direction.

In the diffraction geometry, using the parallel-beam configuration, a 2 $\mu$m pinhole carried by the 2D piezo motors was placed at 2.2 mm downstream the sample in the diffracted beam direction and used for scanning. The smaller, 2 $\mu$m pinhole was used for appropriate sampling for ptychography at 2.171 m. The data was acquired with the Eiger 500K detector, placed at 2.171 m along the diffracted beam direction. Samples were rocked around the (2 0 2) diffraction peak in a range of $\pm$0.02$^{\circ}$ with an angular step of 0.002$^{\circ}$. Ptychography scans were performed with a step size of 0.5 $\mu$m and 0.2 s exposure time following a Fermat spiral pattern \cite{36Huang2014}.

\begin{figure*}
\includegraphics[width=1.\textwidth]{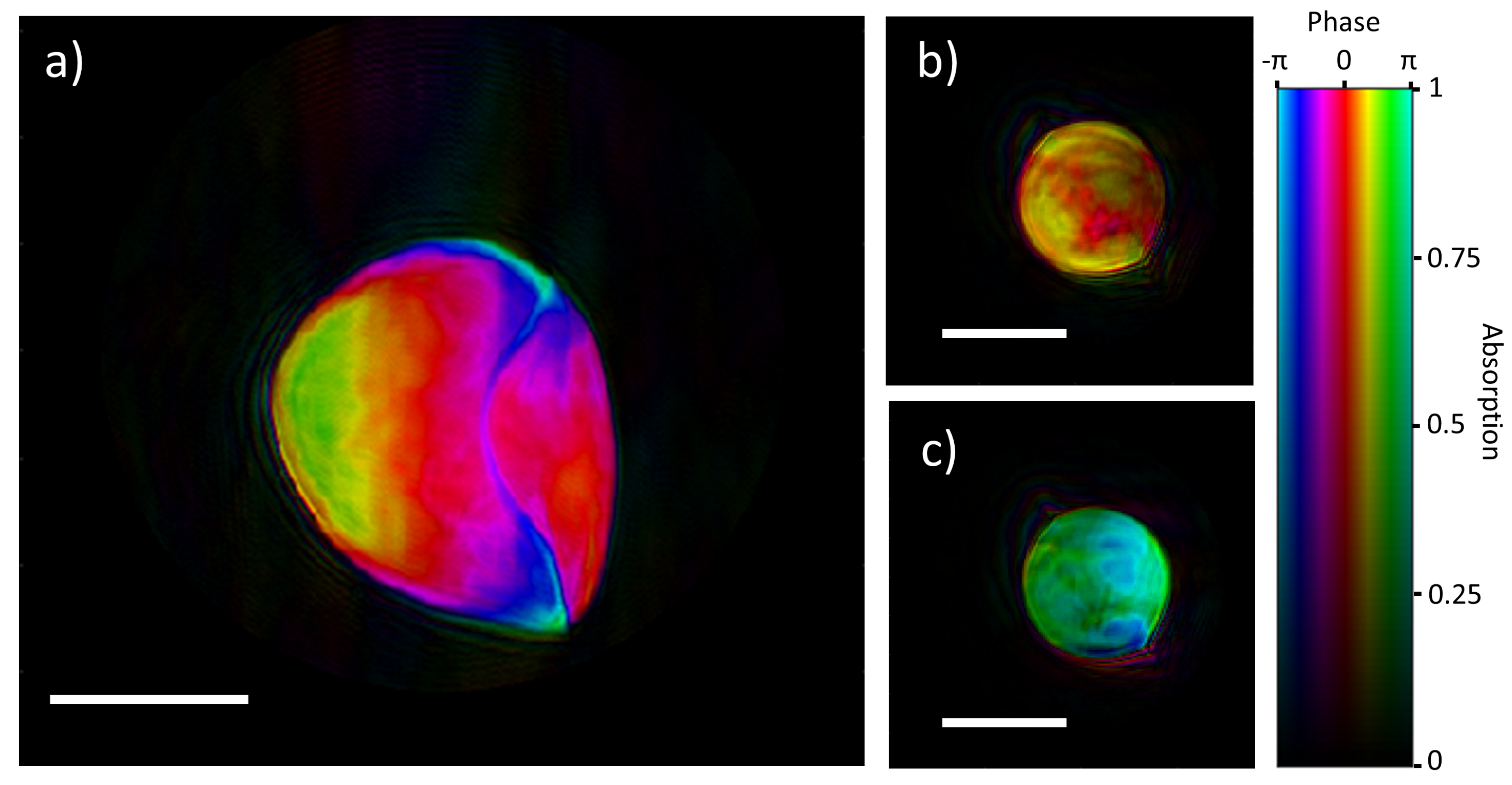}
\caption{\label{fig:sf1}Resulting ptychographic reconstructions of the pinholes: a) in forward geometry, b)  and c) in diffraction geometry, 77\% and 23\% contribution respectively. Scale bar: 2 $\mu$m. Note: phase ramp was removed for clarity of the images. Phase ramp is an intrinsic degree of freedom in ptychographic reconstructions, as discussed in \cite{40Guizar-Sicairos2011}.}
\end{figure*}

\begin{figure*}
\includegraphics[width=0.8\textwidth]{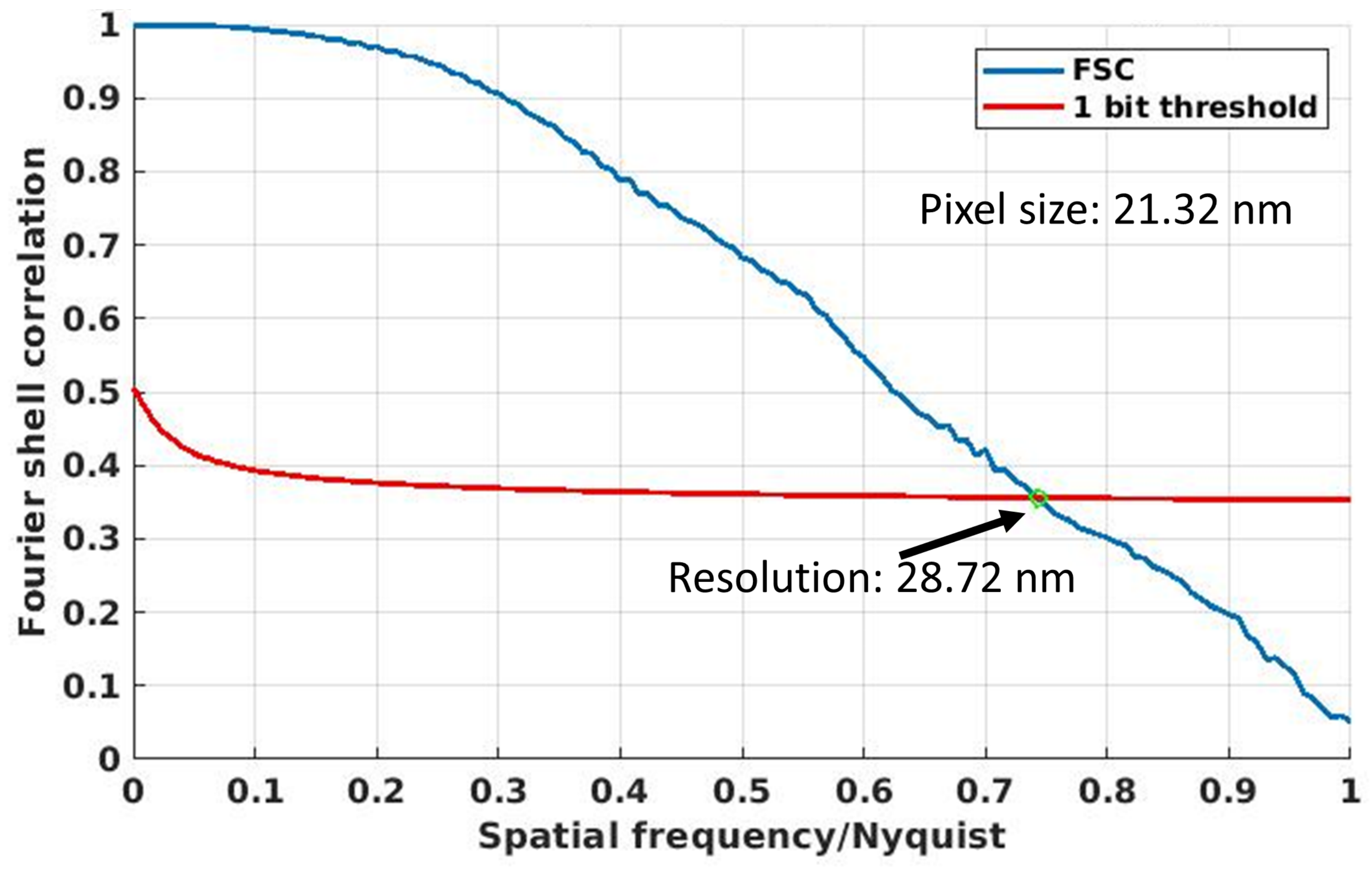}
\caption{\label{fig:sf2}Resolution estimated using Fourier ring correlation of two individually reconstructed repeated scans in forward ptychographic topography. The region taken into account for the estimation comes from within the sample at the bottom of the pillar and beginning of the pedestal, where strain features are observed.}
\end{figure*}

\begin{figure*}
\includegraphics[width=0.9\textwidth]{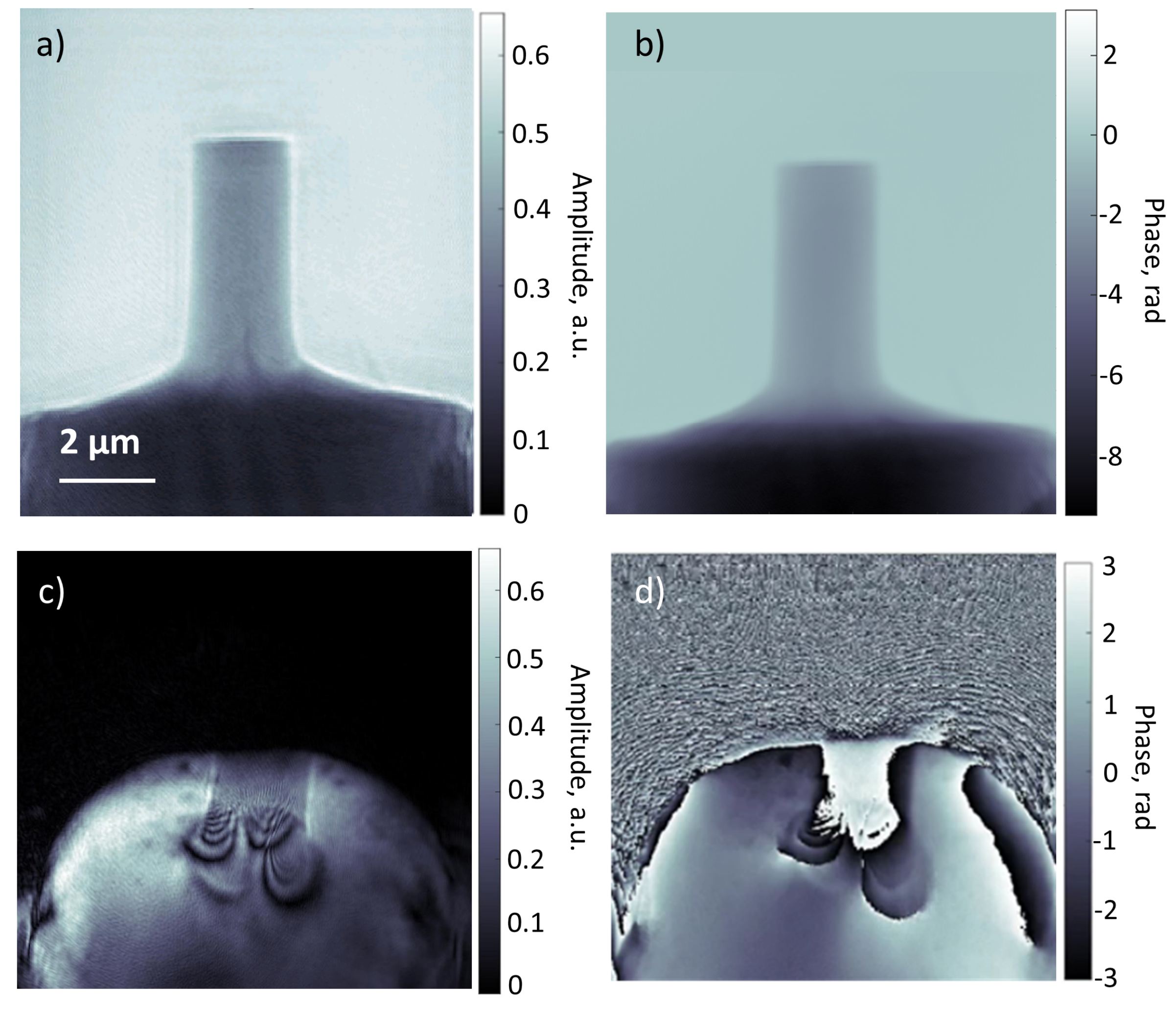}
\caption{\label{fig:sf3}(a) Amplitude and (b) phase of the image reconstruction obtained in forward direction at the rocking angle of 0.03$^{\circ}$. (c) Amplitude and (d) phase of the image reconstruction in diffraction direction at the rocking angle of 0.002$^{\circ}$.}
\end{figure*}

\begin{figure*}
\includegraphics[width=0.9\textwidth]{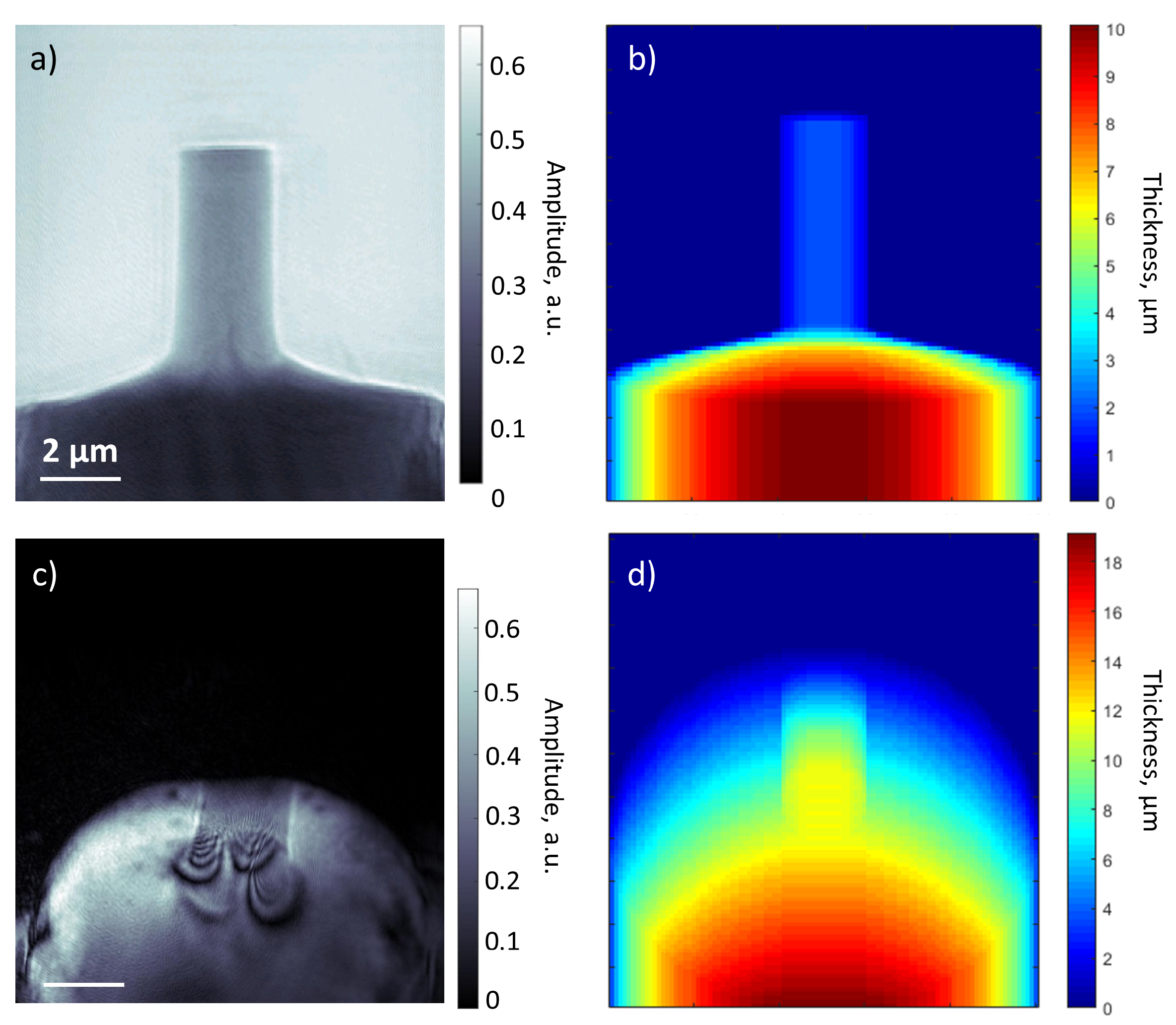}
\caption{\label{fig:sf4}Simulations of the sample thickness along the beam path: (a) amplitude of forward and (c) diffraction direction ptychographic reconstruction, (b) projected sample thickness along the forward direction, (d) projected sample thickness along the diffraction direction.}
\end{figure*}

\begin{figure*}
\includegraphics[width=1.\textwidth]{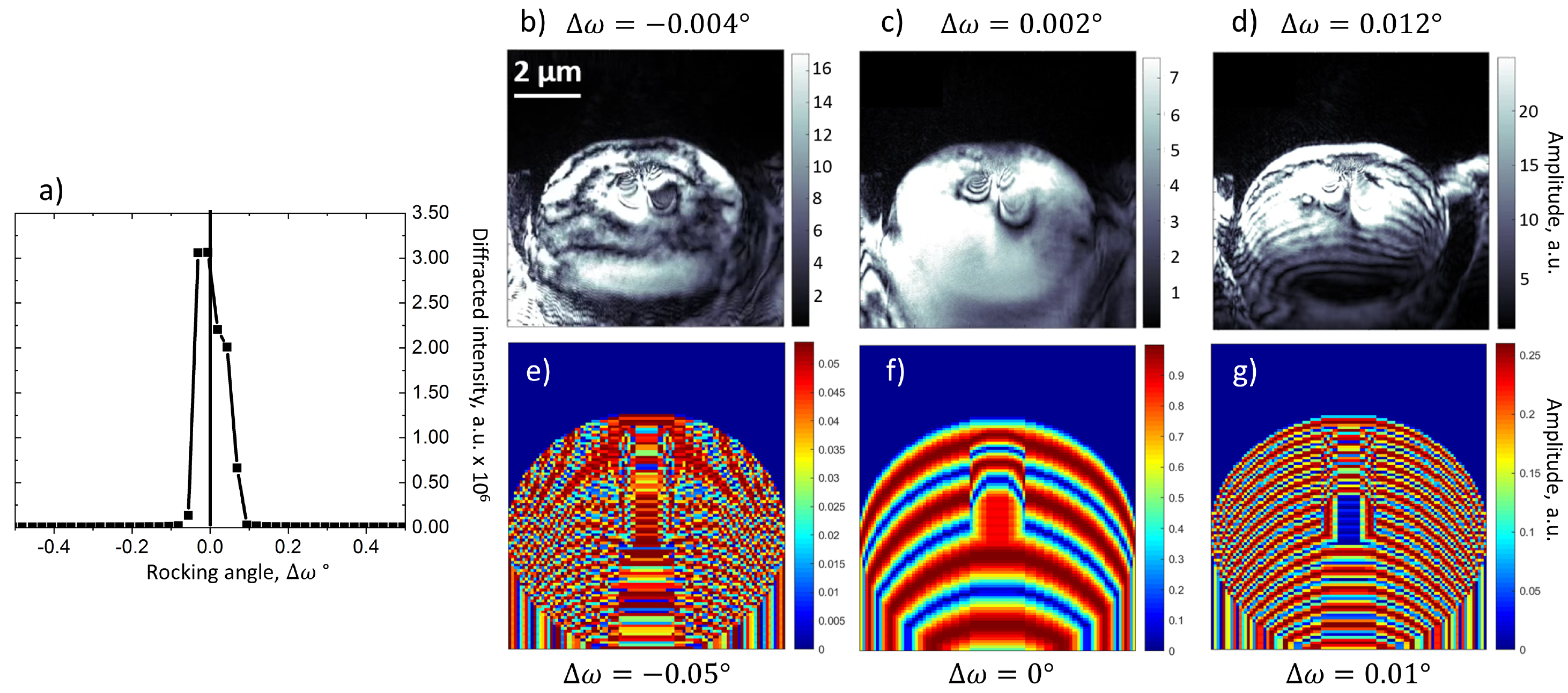}
\caption{\label{fig:sf5}Pendellösung fringes simulations in symmetrical Laue-case with the pencil beam: (a) rocking curve around the (2 0 2) diffraction peak, acquired with the focused beam at the pedestal position. (b)-(d) Amplitudes of the ptychographic topography reconstructions in diffraction direction at specified rocking angles $\bm{\Delta \omega}$. (e)-(g) Simulated amplitudes based on pencil-beam symmetrical Laue dynamical theory, the Eq. 16 from [27] with (f) being at diffraction angle and (e) and (g) being at $\bm{\Delta \omega}$ from the diffraction angle.}
\end{figure*}

\begin{figure*}
\includegraphics[width=1.\textwidth]{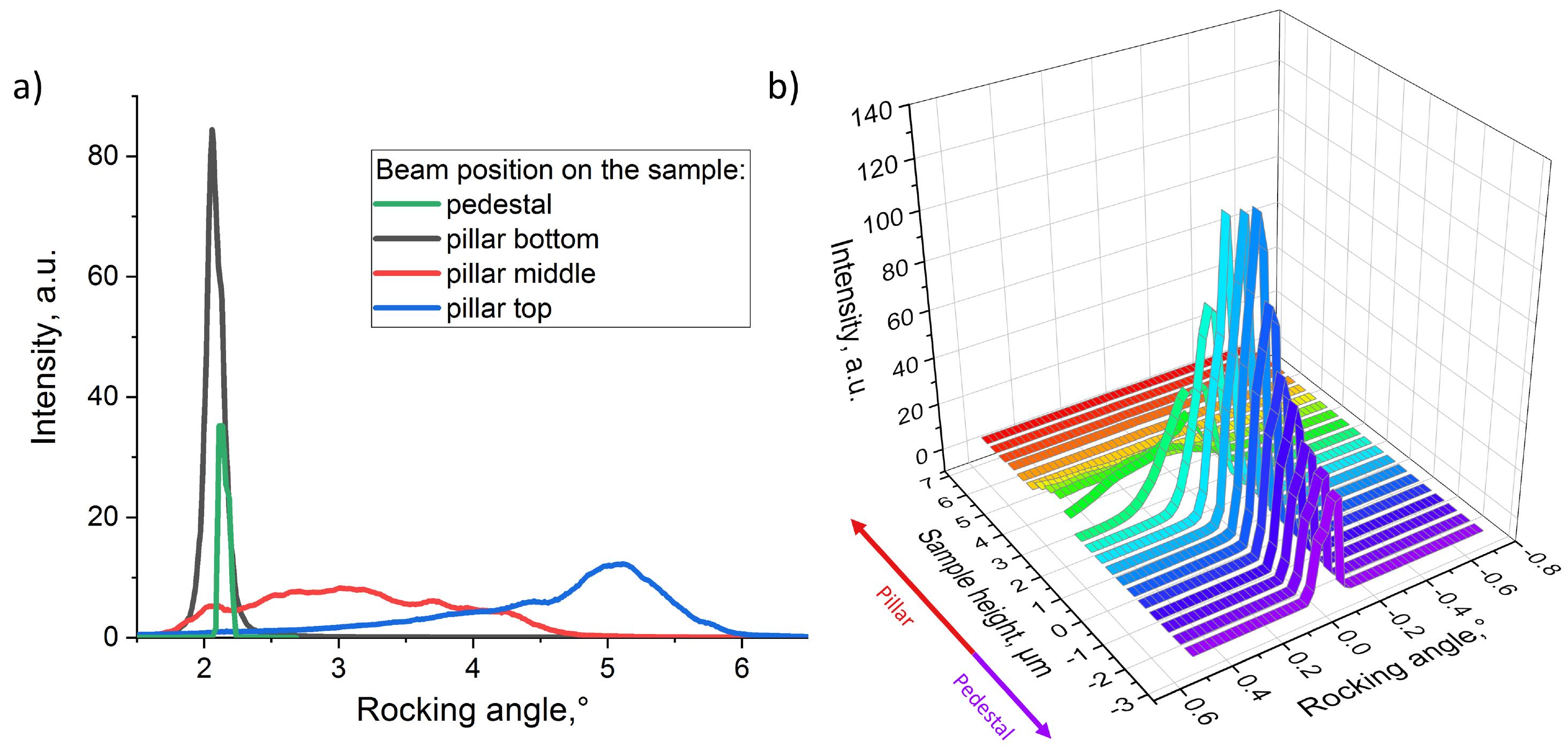}
\caption{\label{fig:sf6}Intensity of the (2 0 2) diffraction peak integrated over all pixels of the detector as a function of rocking angle, acquired with focused beam: (a) comparison of integrated intensity on the diffraction peak at four characteristic positions (at the pedestal, bottom, middle and top of the pillar), (b) the (2 0 2) diffraction peak evolution along the pillar and pedestal height. Position of 0 height correspond to the pillar/pedestal interface.}
\end{figure*}

\begin{figure*}
\includegraphics[width=1.\textwidth]{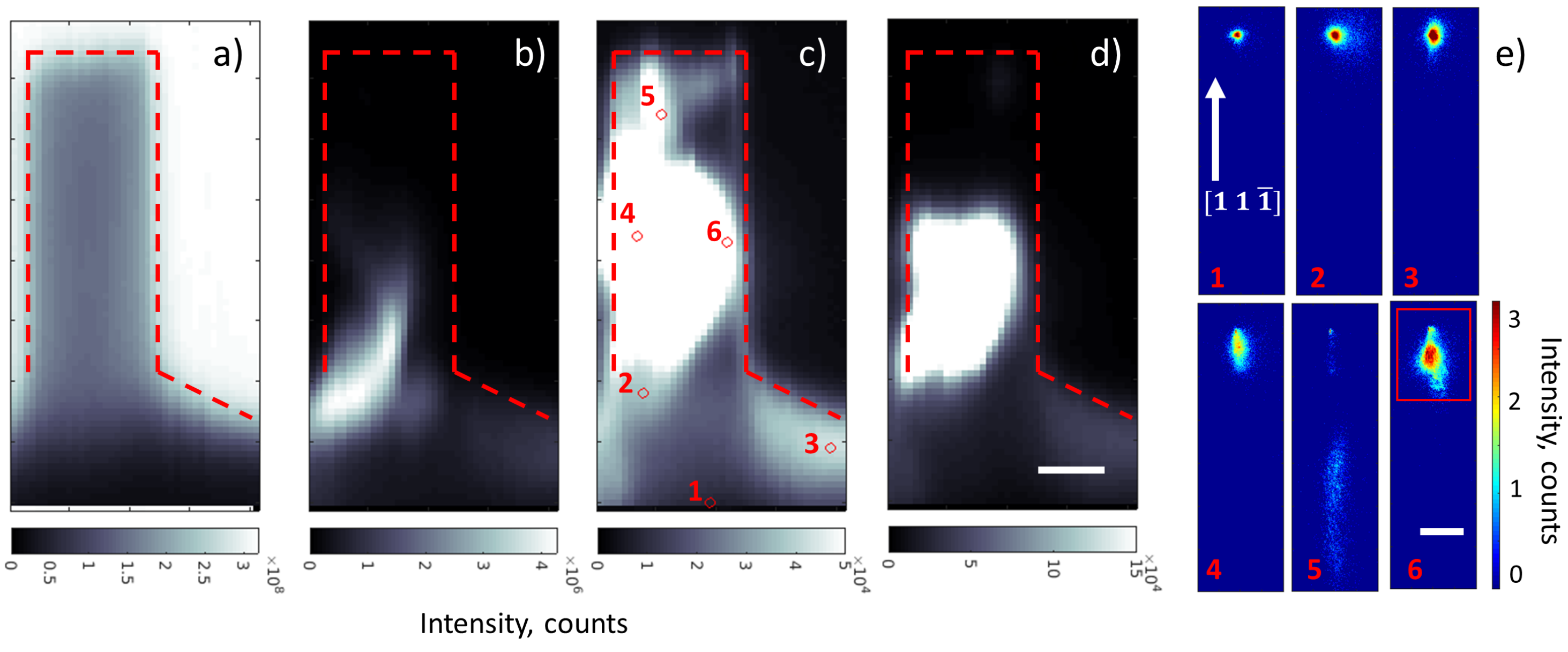}
\caption{\label{fig:sf7}Scanning X-ray microscopy of the S1 sample with the focused beam of 100 nm diameter, scanning step size 100 nm, 0.1 s acquisition time. Sample is scanned at the rocking angle of -0.05$^{\circ}$. (a) Intensity integrated over the central area around the beam of the forward detector. (b) Intensity integrated over the entire area of the detector at diffraction direction. (c) Dark field image obtained by masking out only the peak from the perfect crystal (pedestal position 1) with characteristic diffraction patterns in (e). (d) Dark field image obtained by masking out the peak from the perfect crystal and using the ROI shown in (e) as a red rectangle. This ROI corresponds to the mild deformation, without taking into account very strong deformation in the form of peak splitting. Scale bar in (a)-(d): 1 $\mu$m. Scale bar in (e): 34.55 $\mu$m$^{-1}$.}
\end{figure*}

\begin{figure*}
\includegraphics[width=1.\textwidth]{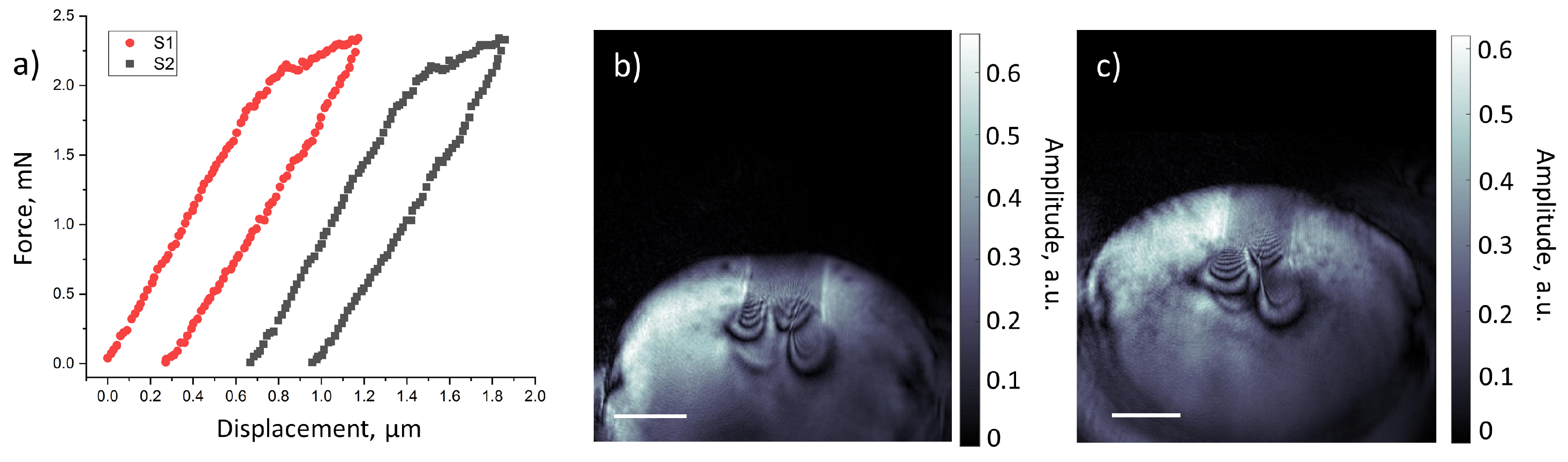}
\caption{\label{figs:f8}Ptychographic topography results on the sample S2 shown for reproducibility purpose: (a) force-displacement curves of the compression test for samples S1 and S2. The mechanical curve for S2 has been artificially shifted for better comparison with the one for S1. Amplitude of the ptychographic topography reconstruction in the diffraction direction for the sample (b)~S1 and (c) S2. Scale bar: 2 $\mu$m.}
\end{figure*}

\begin{figure*}
\includegraphics[width=1.\textwidth]{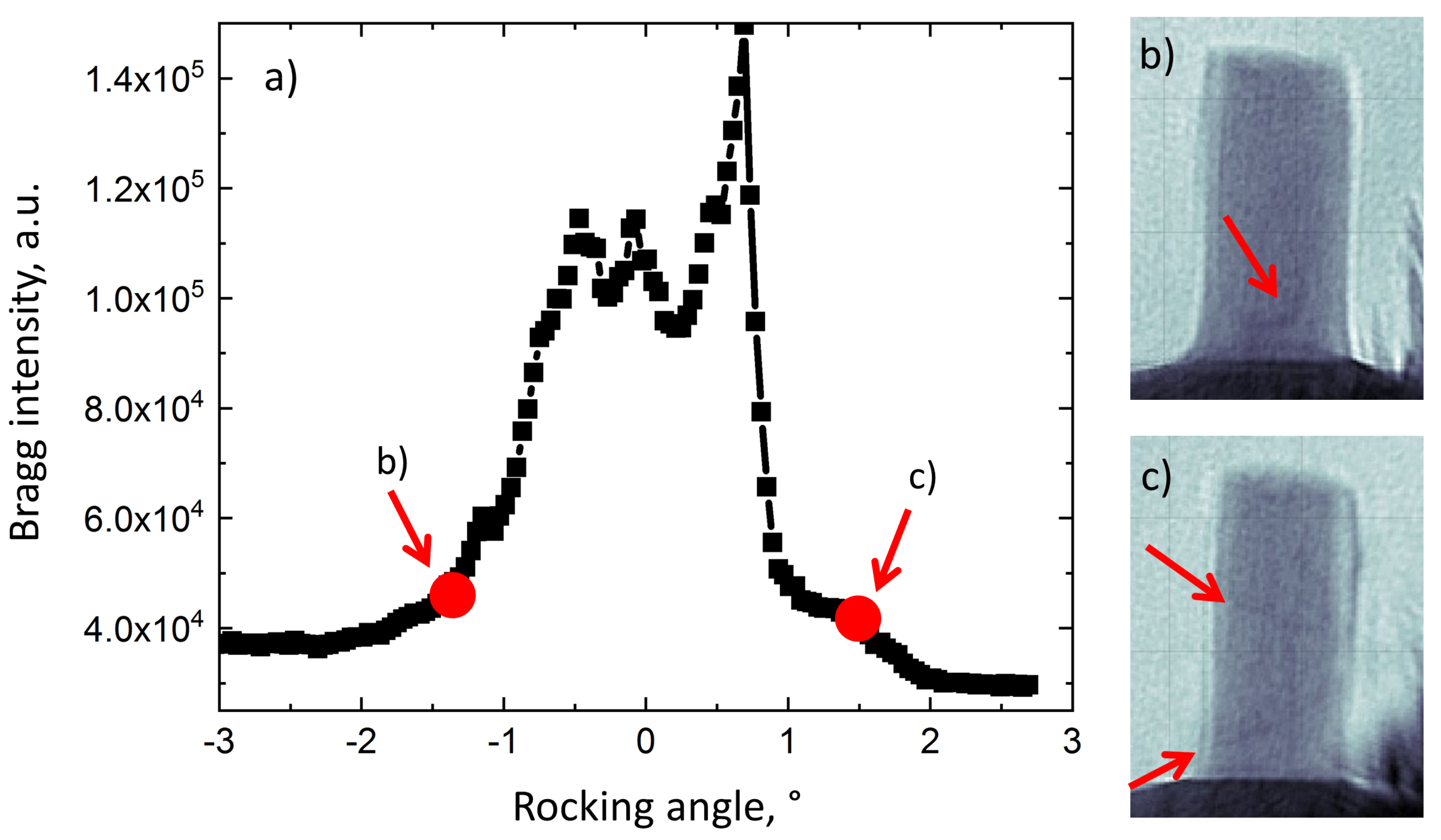}
\caption{\label{fig:sf9}Ptychographic topography on the sample S2 in the forward direction. (a) Rocking curve with the focused beam at the center of the pillar position around (2 2 0) Bragg reflection. (b) and (c) amplitude of the ptychographic topography reconstruction in the forward direction at the right and left shoulder of the Bragg peak. Two features are visible, shown with arrows in (b) and (c). The feature shown with the red arrow in (b) is similar to the one observed in sample S1.}
\end{figure*}

\clearpage

\providecommand{\noopsort}[1]{}\providecommand{\singleletter}[1]{#1}%